\documentclass[journal]{IEEEtran}
\usepackage{latexsym,,amssymb,amsmath,graphicx,epsf,cite,bbm,float}
\usepackage{ifpdf}
\usepackage{epstopdf,mathtools}

\usepackage{algorithm,algorithmic}
\usepackage{amsmath,amssymb,bm}
\usepackage{amsfonts,dsfont,color,bbm,subcaption}

\usepackage{mathtools}

\def\ninept{\def\baselinestretch{1}}
\ninept

\newcommand{\abs}[1]{|#1|}

\DeclareMathOperator{\sign}{sgn}
\DeclareMathOperator*{\argmax}{arg\,max}
\DeclareMathOperator*{\argmin}{arg\,min}

\usepackage{hyperref,amsthm}

\newtheorem{theorem}{Theorem}

\newtheorem{lemma}[]{Lemma}

\newtheorem{proposition}[]{Proposition}

\newtheorem{remark}[]{Remark}

\newtheorem{definition}[]{Definition}

\newtheorem{assumption}[]{Assumption}

\newtheorem{example}[]{Example}

\begin{document}

\title{Efficient Locally Optimal Number Set Partitioning for Scheduling, Allocation and Fair Selection} 
\author{\IEEEauthorblockN{Kaan Gokcesu}, \IEEEauthorblockN{Hakan Gokcesu} }
\maketitle

\begin{abstract}
	We study the optimization version of the set partition problem (where the difference between the partition sums are minimized), which has numerous applications in decision theory literature. While the set partitioning problem is NP-hard and requires exponential complexity to solve (i.e., intractable); we formulate a weaker version of this NP-hard problem, where the goal is to find a locally optimal solution. We show that our proposed algorithms can find a locally optimal solution in near linear time. Our algorithms require neither positive nor integer elements in the input set, hence, they are more widely applicable.
\end{abstract}

\section{Introduction}\label{sec:intro}
\subsection{Set Partition Problem}
In the traditional set partition problem \cite{cook1971complexity,turing1937computable,levin1973universal,korf1998complete}, the goal is to decide whether a given input set $\mathcal{X}$ of positive integers can be partitioned into two disjoint complementary subsets $\mathcal{X}_1$ and $\mathcal{X}_2$ such that the sum of the elements in $\mathcal{X}_1$ equals to the sum of the elements in $\mathcal{X}_2$. A closely related problem to the set partitioning is the subset-sum problem \cite{kleinberg2006algorithm}, where the goal is to find a subset, sum of which equals a target value $T$. The subset-sum problem can be solved with a set partition solver by adding a dummy sample. In this work, we deal with the optimization version of the partition problem, which is to partition the set $\mathcal{X}$ into two subsets $\mathcal{X}_1$ and $\mathcal{X}_2$ such that the absolute difference between the sum of elements in $\mathcal{X}_1$ and the sum of elements in $\mathcal{X}_2$ is minimized. 

The set partition problem is one of Karp's $21$ NP-complete combinatorial problems \cite{karp1972reducibility}. It is also one of Garey and Johnson’s $6$ NP-complete fundamental problems \cite{garey1979computers}.
Similarly, the subset sum problem is also NP-complete \cite{garey1979computers}; and the optimization version, which we investigate, is NP-hard \cite{korf2009multi}. For the partition problem, despite the NP-completeness, there exist heuristics that can solve it optimally or approximately in many instances. Moreover, there are situations, where an exponential number of optimal solutions exist. Therefore, it has also been called "the easiest hard problem" in literature because of its limited structure in comparison to other NP-complete problems \cite{hayes2002computing,mertens2006number}. 
Although the NP-completeness imply an exponential complexity solver in general, there exists pseudo-polynomial time algorithms for the set partition problem, where the run time of the algorithm is polynomial in the numeric value of the inputs (in contrast to polynomial time which is polynomial in the length of the input) \cite{garey1979computers}. 
However, since it is NP-complete, these algorithms have limited uses, i.e, they are intractable for high precision input.
Nevertheless, it is a well-studied problem with continuous algorithmic improvements over time \cite{graham1966bounds,coffman1978application,dell1995optimal,korf1998complete,korf2009multi,moffitt2013search,schreiber2013improved,schreiber2014cached,schreiber2018optimal}, which leads to the possibility that similar improvements may be probable for harder NP-complete problems \cite{schreiber2018optimal}. 

\subsection{Applications of Set Partitioning}
The set partitioning problem has a lot of applications in learning, optimization and decision problems, e.g., scheduling, encryption, allocation, classification and training data splits \cite{neyshabouri2018asymptotically,cano2006combination,cano2007evolutionary,gokcesu2018adaptive,garcia2009enhancing,gokcesu2020generalized,golbraikh2000predictive,fan2005working,gokcesu2020recursive,huberbook,cesabook,poor_book,furusjo2006importance,gokcesu2021generalized,sarkar1987partitioning,dell2008heuristic,graham1979optimization,gokcesu2021optimal,walsh2009really,merkle1978hiding,shamir1982polynomial,rivest1983cryptographic,gokcesu2021optimally,buhrkal2011models,umang2013exact,iris2015integrated,lalla2016set,gokcesu2021regret}. Many interesting real life scenarios are as follows.
\subsubsection{Processor Scheduling}
In some literature, the set partition problem has become synonymous with the processor scheduling problem \cite{garey1979computers,sarkar1987partitioning,dell2008heuristic,graham1979optimization}. The input set $\mathcal{X}$ of $N$ positive elements correspond to the run times of a set of $N$ jobs. The goal of the processor scheduling is to assign each of the $N$ jobs to one of working identical machines (such as processor cores that execute in parallel), while minimizing the time it takes to complete all the jobs in the schedule. Minimizing the total schedule time is equivalent to the minimization of the schedule time on the machine with the maximum load.

\subsubsection{Berth Allocation}
Similarly, the berth allocation problem (also known as the berth scheduling problem) \cite{buhrkal2011models,umang2013exact} is an NP-complete problem in the field of operations research, which deals with the allocation of berth space for vessels in container terminals. In this problem, the vessels arrive over time and the terminal operator needs to assign them to berths for loading or unloading of containers as soon as possible. Different factors may affect the individual berth and time assignment of the vessels. The set partition methods can be straightforwardly utilized in this situation to minimize the time it takes to serve all vessels \cite{iris2015integrated,lalla2016set}.

\subsubsection{Fair Team Selection}
In schoolyards, the traditional method of choosing fair teams is to assign one captain (selector) for each team and then have each captain pick players in round-robin fashion. Let each element in the input set correspond to a player's skill level, and let the strength of a team be equal to the sum of the player skills. The traditional method of picking the best player remaining does not necessarily lead to fair teams, and an optimal set partition could lead to a better team selection \cite{hayes2002computing}.

\subsubsection{Voting Manipulation}
Suppose we have a veto election, where instead of voting for a candidate, voters veto a candidate (with each voter’s veto carrying a different weight) \cite{walsh2009really}. If the candidate with the smallest total veto wins the election, a manipulative group's best strategy will be to partition their veto weights among the opposing candidates using a set partition algorithm. Only then can the veto weights of these candidates be larger than the veto weight of their own candidate.

\subsubsection{Public Key Encryption}
Merkle-Hellman knapsack cryptosystem \cite{merkle1978hiding} is an early encryption method, which is based on the subset-sum problem (hence, the set partition problem). While polynomial time algorithms have been found for their most basic encryption \cite{shamir1982polynomial}, it was a pioneer work towards more powerful encryption systems such as RSA \cite{rivest1983cryptographic}.

\subsection{Literature Review}
Even though the set partition problem is NP-hard. There have been approximation algorithms developed to solve the set partition problem approximately in polynomial complexity. 
\subsubsection{Greedy Algorithm}
The most obvious method for the set partition problem is the greedy algorithm \cite{graham1966bounds}. The method first sorts the input set $\mathcal{X}$ in decreasing order and considers the elements one by one to place them into the subset with the smallest sum so far. Whenever the sums are equal, it selects one of the subsets arbitrarily. 

The greedy algorithm runs in $O(N\log N)$ time and $O(N)$ space. The partition values are within $4/3$ of optimal \cite{kellerer2004knapsack}. The greedy algorithm is optimal for $N\leq 4$ \cite{korf2011hybrid}.

\subsubsection{Set Differencing: The Karmarkar-Karp Algorithm (KK)}
The Karmarkar-Karp (KK) set differencing algorithm \cite{karmarkar1982differencing} is an alternative to the greedy approach. Similarly, it begins by sorting $\mathcal{X}$ in decreasing order. After that, it replaces the largest two elements of $\mathcal{X}$ with their difference in an iterative fashion. This is the same as placing the two largest integers into different subsets without specifying which integer goes to which. It iteratively continues in this manner by replacing the two largest elements with their difference until there is only one element left, which is the difference between the subset sums. To reconstruct the two subsets, KK backtracks the actions taken in its iterations, and generates the partition by performing them in reverse. 

The KK algorithm runs in $O(N\log N)$ time and $O(N)$ space. Although it is better in practice than the greedy algorithm, it still produces a suboptimal solution.

\subsubsection{Dynamic Programming (DP)}
A dynamic programming (DP) approach \cite{garey1979computers,martello1990knapsack,korf2013optimally} solves the partition problem in pseudo-polynomial time and space. Let $S$ be the sum of all elements of the input set $\mathcal{X}$ (which consists of all positive integers). DP creates a binary matrix $M$ with $N$ rows and $S+1$ columns (where $N$ is the cardinality of $\mathcal{X}$). The rows correspond to the integers in $\mathcal{X}$ in descending order, and the columns correspond to the possible sums in $\{0,1,2,\ldots,S\}$. A bit at row $r$ and column $c$ is set to one if there is a subset of the largest $r$ integers of $\mathcal{X}$ which sum to $c$. To create this matrix, DP starts as follows. At the beginning, it sorts $\mathcal{X}$ in descending order, sets all bits in $M$ to zero except for the $0^{th}$ column which is set to one. Then, it sets the column corresponding to the largest integer of $\mathcal{X}$ in the first row to one. For each row $r$, it copies all of the one-bits in the previous row $r-1$ to the current row. Then, for each column $c$ in the previous row that has a bit set to one, it sets column $c+x_r$ in the current row to one, where $x_r$ is the integer in $\mathcal{X}$ corresponding to the current row $r$. The one-bits in the last row correspond to all subset sums that can be formed from $\mathcal{X}$. The smallest sum greater than or equal to $S/2$ corresponds to the optimal partition. Since each row is created from the preceding row, DP stores just one row, i.e., after creating an entire row, instead of copying, it continues working on that same row. 

DP runs in $O(NS)$ time and $O(S)$ space (hence, pseudo-polynomial). However, its performance is highly dependent on the precision of the input elements. To this end, although it is useful in certain input sets, more encompassing approaches are needed.

\subsubsection{The Horowitz-Sahni Algorithm (HS)}
The Horowitz-Sahni (HS) algorithm \cite{horowitz1974computing} starts by calculating an upper bound $S_u$ (possibly using any sub-optimal approach or simply $S$), a corresponding lower bound $S_l=S-S_u$ (i.e., its complement) and the perfect partition value $S_*=S/2$. HS sorts the input set $\mathcal{X}$ into decreasing order and divides it into two sets $\mathcal{X}_A$ and $\mathcal{X}_B$ of size $N/2$ each. It then generates the lists $A$ and $B$ of the sums of all $2^{N/2}$ subsets of each half set. The sums are sorted such that $A$ and $B$ are in ascending and descending order respectively. It then iterates through each element $a$ in $A$ and $b$ in $B$ starting with the first ones. If $a+b$ is less than the lower bound $S_l$, it gets the next $a$ from $A$. If $a+b$ is between $S_l$ and $S_*$, it gets the next $a$ from $A$ after setting $S_l$ to $a+b$ and $S_u$ to its complement $S-a-b$. If $a+b$ is $S_*$, it is returned and HS stops. If $a+b$ is between $S_*$ and $S_u$, it gets the next $b$ from $B$ after setting $S_u$ to $a+b$ and $S_l$ to its complement $S-a-b$. If $a+b$ is greater than $S_u$, it gets the next $b$ from $B$. This iteration continues until there is no more elements in either $A$ or $B$.

$A$ and $B$ takes $O(2^{N/2})$ time to create since they are the power sets of $\mathcal{X}_A$ and $\mathcal{X}_B$ respectively, each of which contains $N/2$ elements. Sorting them takes $O(N2^{N/2})$ time each. Since $A$ and $B$ are scanned in linear time in their sizes, HS runs in time $O(N2^{N/2})$ and $O(2^{N/2})$ space, which is a clear improvement over the brute force complexity $O(2^N)$.

\subsubsection{The Schroeppel-Shamir Algorithm (SS)}
The Schroeppel-Shamir (SS) algorithm \cite{schroeppel1981t} is based on HS with a more efficient use of memory. While HS creates the entire $A$ and $B$ lists and stores them in memory before scanning, SS generates the subset sums of $A$ and $B$ on demand. SS starts by dividing $\mathcal{X}$ into four sets $\mathcal{X}_{A1}$, $\mathcal{X}_{A2}$, $\mathcal{X}_{B1}$, $\mathcal{X}_{B2}$ of sizes $N/4$ each. Then, it creates the lists $A1$, $A2$, $B1$ and $B2$ of all $O(2^{N/4})$ subsets of each quarter set sorted by their subset sums in increasing order. The subsets from $A1$ and $A2$ lists are combined in a min heap to create the subsets in the same order as the list $A$ in HS. Each subset of the heap consists of one subset from each of the $A1$ and $A2$ lists. Initially, it contains all pairs combining the empty set from the $A1$ list with each subset from the $A2$ list. The top of the heap contains the pair with smallest subset sum. Whenever a pair $(A1[i], A2[j])$ is popped off the top of the heap, it is replaced in the heap by a new pair $(A1[i + 1], A2[j])$. Similarly, the subsets from the $B1$ and $B2$ lists are combined in a max heap, which creates subsets in the same order as the $B$ list from HS. SS uses these heaps to generate the subset sums in a sorted order and scans them in the same manner as the HS algorithm.

Generating the $A1$, $A2$, $B1$ and $B2$ lists takes $O(2^{N/4})$ time since they are the power sets of $\mathcal{X}_{A1}$, $\mathcal{X}_{A2}$, $\mathcal{X}_{B1}$, $\mathcal{X}_{B2}$ each of which contains $N/4$ elements. Sorting $A1$, $A2$, $B1$ and $B2$ takes $O(N2^{N/4})$ time each. Scanning the lists will generate the same subsets as HS. Therefore, the scanning operation has time complexity $O(N2^{N/2})$. The overall time complexity is $O(N2^{N/2})$, which is the same time complexity as HS. However, SS only requires $O(2^{N/4})$ space for the four quarter sets and heaps, which is a clear improvement over $O(2^{N/2})$ space complexity of HS.

\subsubsection{Complete Anytime Algorithm (CAA)}
In spite of the algorithmic developments, an optimal solver still requires exponential complexity time and space. Although the exponential complexity is not possible to remove in general because of the NP-hardness, the memory improvement can be made by use of a hierarchical approach. One such method is the complete anytime algorithm (CAA) \cite{korf1998complete}, which can transform the efficient sub-optimal algorithms into optimal ones. Like them, CAA first sorts $\mathcal{X}$ into decreasing order and then proceeds to partition $\mathcal{X}$ into two subsets $\mathcal{X}_1$ and $\mathcal{X}_2$. This partitioning is done by searching a binary tree, where each branch is created according to a decision rule or base algorithm. When the greedy algorithm is used as a base; the left branch puts the element into the subset with the smaller sum (i.e., the greedy choice), and the right into the other subset. At anytime, CAA returns the smallest maximum subset sum encountered so far at the leaf nodes of the binary tree. When the KK algorithm is used as a base; the left branch puts the two largest remaining integers into different subsets by replacing them with their difference (i.e., the KK choice), and the right puts them into the same subset by replacing them with their sum. At anytime, CAA keeps track of the leaf values and returns the smallest absolute leaf value encountered so far, which is the difference between the two subset sums given the path to the leaf. 

The binary tree is searched depth first from left to right. The worst-case time complexity is $O(2^N)$ and the space complexity is $O(N)$, where $N$ is the size of $\mathcal{X}$. Even though it decreases the space complexity considerably, it still has exponential worst-case time complexity. We point out that when the branch creation rule is the direct assignments to the subsets, it becomes the brute-force approach.

\subsection{Contributions and Organization}
Although the greedy algorithm and KK have $O(N\log N)$ time complexities, they only generate sub-optimal solutions. While DP finds an optimal one with pseudo-polynomial complexity $O(NS)$, it has limited use for high precision or non-integer inputs. Even though HS and SS are able to find optimal solutions regardless, they have exponential complexities (albeit better than brute force). CAA improves upon them to find an optimal solution with linear $O(N)$ memory. While it incorporates the sub-optimal methods in its algorithm for better performance, its time complexity is the same as brute force in the worst case. Improving the exponential complexity is unfruitful because of the NP-hardness of the problem but fast algorithms are always desired especially with the emergence of big data. To this end, we tackle the problem from a different point of view by trying to solve a 'weaker' version. We postulate that finding a 'locally' optimal solution to the set partition problem is nowhere near as hard as finding a 'globally' optimal one. In \autoref{sec:prob}, we mathematically formulate the problem definition. In \autoref{sec:method}, we provide an efficient algorithm that can find a locally optimal solution in $O(N\log N)$ time and $O(N)$ space. In \autoref{sec:method2}, we improve upon our design for better performance. We extend our methods to any real inputs in \autoref{sec:ext} and finish with concluding remarks in \autoref{sec:conc}.

\section{Locally Optimal Set Partition Problem}\label{sec:prob}
In this section, we formally define the set partition problem. As an input, we have the set of numbers
	\begin{align}
	\mathcal{X}=&\{x_1,x_2,\ldots,x_N\},\\
	=&\{x_n\}_{n=1}^N
	\end{align}	
We partition $\mathcal{X}$ into two subsets $\mathcal{X}_1$ and $\mathcal{X}_2$ such that they are disjoint and their union is $\mathcal{X}$, i.e.,
	\begin{align}
	\mathcal{X}_1\cap\mathcal{X}_2&=\emptyset,\\
	\mathcal{X}_1\cup\mathcal{X}_2&=\mathcal{X},
	\end{align}	
The goal is to create the sets $\mathcal{X}_1$ and $\mathcal{X}_2$ such that their individual sums are as close as possible to each other. Hence, in this problem, we compare the set sums with each other, which are denoted as
\begin{align}
S_1=\sum_{x\in\mathcal{X}_1}x,\\
S_2=\sum_{x\in\mathcal{X}_2}x,
\end{align}
Hence, $S_1$ and $S_2$ are each other's complements, i.e., $S_1=S-S_2$ and $S_2=S-S_1$, where
\begin{align}
	S=\sum_{x\in\mathcal{X}}x.
\end{align}
We consider the formulation where we want to minimize the absolute difference between the sums $S_1$ and $S_2$, i.e.,
\begin{align}
\min_{\mathcal{X}_1,\mathcal{X}_2} \left(\abs{S_1-S_2}\right).
\end{align}
However, we point out that the formulation of minimizing the maximum sum, i.e.,
\begin{align}
\min_{\mathcal{X}_1,\mathcal{X}_2} \left(\max(S_1,S_2)\right),
\end{align}
or the formulation of maximizing the minimum sum, i.e.,
\begin{align}
\max_{\mathcal{X}_1,\mathcal{X}_2} \left(\min(S_1,S_2)\right),
\end{align}
are all equivalent. Since we are only dealing with two sets and their corresponding sums, in general, all such formulations become equivalent to each other.

This problem is unfortunately NP-hard \cite{karp1972reducibility}, and is impossible to solve with an efficient method. To this end, instead of this NP-hard problem, we consider a 'weaker' version of the set partition problem. Instead of a global optimal solution, we are after a 'locally' optimal one, which is defined as follows:

\begin{definition}
	A partition $\mathcal{X}_1$ and $\mathcal{X}_2$ of $\mathcal{X}$ is locally optimal if there is no single element transfer that can decrease the absolute difference between the set sums $S_1$ and $S_2$, i.e.,
	\begin{align}
		\abs{(S_1-x)-(S_2+x)}\geq&\abs{S_1-S_2}, &&\forall x\in\mathcal{X}_1,\\
		\abs{(S_1+x)-(S_2-x)}\geq&\abs{S_1-S_2}, &&\forall x\in\mathcal{X}_2,
	\end{align}
\end{definition}

Next, we provide an efficient procedure which can find such a locally optimal partition.

\section{Methodology}\label{sec:method}
\subsection{Iterative Algorithm}\label{sec:alg}
Before we propose the algorithm, we make some initial assumptions.
\begin{assumption}
	Let the set $\mathcal{X}$ be composed of only positive elements, i.e.,
	\begin{align*}
		x>0, &&\forall x\in \mathcal{X}.
	\end{align*}
\end{assumption}
\begin{assumption}
	Let the set $\mathcal{X}=\{x_n\}_{n=1}^N$ be in ascending order, i.e.,
	\begin{align*}
	x_n\leq x_{n+1}, &&\forall n\in\{1,2,\ldots,N-1\}.
	\end{align*}
\end{assumption}
Thus, our input $\mathcal{X}$ is a positive ordered set. However, we point out that there is no requirement for the elements to be integers. Given the input $\mathcal{X}$ set, the algorithm works as follows: 
\begin{enumerate}
	\item Create the set $\widetilde{\mathcal{X}}=\{\tilde{x}_n\}_{n=1}^N$, where $\tilde{x}_n=x_n$.
	\item\label{step:max} Find the maximum $\tilde{x}$ that is strictly less $S_1-S_2$, i.e., 
	\begin{align}
	\tilde{x}_*=\argmax_{\tilde{x}\in\widetilde{\mathcal{X}}}\tilde{x}: &&\tilde{x}<S_1-S_2.\label{max}
	\end{align}
	\item Then, move the element $\tilde{x}_*$ from $\mathcal{X}_1$ to $\mathcal{X}_2$, and update $S_1$ and $S_2$ accordingly, i.e.,
	\begin{align}
	S_1\leftarrow S_1-\tilde{x}_*,\label{S1}\\
	S_2\leftarrow S_2+\tilde{x}_*.\label{S2}
	\end{align}
	\item Then, change the sign of $\tilde{x}_*$ to represent the transfer
	\begin{align}
	\tilde{x}_*\leftarrow-\tilde{x}_*.\label{x*}
	\end{align}
	\item We stop if $S_1-S_2\leq0\label{S1<S2}$, or else, we return to Step \ref{step:max}.
\end{enumerate}

\begin{remark}
	A few remarks about the algorithm:
	\begin{itemize}
		\item At the beginning of our algorithm, all elements of $\mathcal{X}$ are assigned to $\mathcal{X}_1$, i.e.,
		\begin{align}
			\mathcal{X}_1=\mathcal{X}, && \mathcal{X}_2=\emptyset.
		\end{align}
		\item At the beginning, we have the set sums 
		\begin{align}
			S_1=\sum_{x\in\mathcal{X}}x, &&S_2=0.
		\end{align} 
		\item At each iteration of the algorithm, the positive $\tilde{x}$ will be in $\mathcal{X}_1$ and the negative $x$ will be in $\mathcal{X}_2$. 
		\item At each iteration, one positive $\tilde{x}\in\widetilde{\mathcal{X}}$ has its sign changed to negative.
		\item At each iteration, the difference $S_1-S_2$ decreases.
	\end{itemize}
\end{remark}

\begin{example}
	A working example is as follows. Let the input be $\mathcal{X}=\{2,3,5,7,11,13,17,19,23,29\}$. We sequentially have:
	\begin{align*}
		\widetilde{\mathcal{X}}=&\{2,3,5,7,11,13,17,19,23,29\}, &&S_1-S_2=129,\\
		\widetilde{\mathcal{X}}=&\{-29,2,3,5,7,11,13,17,19,23\}, &&S_1-S_2=71,\\
		\widetilde{\mathcal{X}}=&\{-29,-23,2,3,5,7,11,13,17,19\}, &&S_1-S_2=25,\\
		\widetilde{\mathcal{X}}=&\{-29,-23,-19,2,3,5,7,11,13,17\}, &&S_1-S_2=-13.
	\end{align*}
	Hence, the resulting partition is
	\begin{align*}
		\mathcal{X}_1=\{2,3,5,7,11,13,17\} &&\mathcal{X}_2=\{19,23,29\},
	\end{align*} 
	which is locally optimal.
\end{example}
 
\subsection{Local Optimality}
In this section, we prove the local optimality of our algorithm, where we start with a few useful lemmas.
\begin{lemma}\label{thm:S2-S1<x}
	At the end (stop), we have $0\leq S_2-S_1<\abs{\tilde{x}_*}$.
	\begin{proof}
		Before the final update, we have from \eqref{max}
		\begin{align}
			\tilde{x}_*^{old}<S_1^{old}-S_2^{old}.
		\end{align}
		and combining with Step \ref{S1<S2}, we have
		\begin{align}
			-\tilde{x}_*^{old}<S_1^{old}-S_2^{old}-2\tilde{x}_*^{old}\leq 0.
		\end{align}
		Thus, the absolute difference between the new set sums is bounded as
		\begin{align}
			\abs{S_1^{old}-\tilde{x}_*^{old}-S_2^{old}+\tilde{x}_*^{old}}=&(S_2^{old}+\tilde{x}_*^{old})-(S_1^{old}-\tilde{x}_*^{old}),\\
			<&\tilde{x}_*^{old},
		\end{align}
		which, together with \eqref{S1}, \eqref{S2} and \eqref{x*}, concludes the proof.
	\end{proof}
\end{lemma}

\begin{lemma}\label{thm:x*dec}
	Each element $\tilde{x}$ that we change from positive to negative at every round of the algorithm is in nonincreasing order.
	\begin{proof}
		At each step of the algorithm, we change the sign of the maximum element that is less than the sum difference from \eqref{max}. After that, either the algorithm stops or the sum difference get smaller. Thus, in the subsequent rounds, each element that we make negative will be smaller than the previously switched elements.
	\end{proof}
\end{lemma}

\begin{theorem}
	At the stop, we have a locally optimal set partition solution.
	\begin{proof}
		From \autoref{thm:S2-S1<x}, we have
		\begin{align}
			0\leq S_2-S_1<\abs{\tilde{x}_*}.
		\end{align}
		The only way we can decrease the absolute difference $\abs{S_1-S_2}$ is by moving an element from $\mathcal{X}_2$ to $\mathcal{X}_1$, i.e., changing the sign of a negative element $\tilde{x}$ such that
		\begin{align}
			0>\tilde{x}>-\abs{S_1-S_2}>-\abs{\tilde{x}_*}.
		\end{align}
		However, from \autoref*{thm:x*dec}, we see that there is no such element since all the predecessors of $\tilde{x}_*$ are bigger, which concludes the proof.
	\end{proof}
\end{theorem}

\subsection{Complexity Analysis}

We observe that we iterate through the ordered set $\mathcal{X}$ sequentially in descending order and the algorithm stops at some sample $x\in\mathcal{X}$. Thus, at the worst case, our computational complexity is $O(N)$. If the samples were unordered, we can sort them in $O(N\log N)$ time.
Since we only keep track of the set $\widetilde{\mathcal{X}}$ and update the signs of its elements, our memory complexity is $O(N)$. Thus, the algorithm runs in $O(N\log N)$ time and $O(N)$ space.

\section{Improved Iterative Algorithm}\label{sec:method2}
In this section, we propose an alternative method, which may be globally optimal for certain examples without loss of efficiency. 
We restructure the algorithm as follows: 
\begin{enumerate}
	\item Create the set $\widetilde{\mathcal{X}}=\{\tilde{x}_n\}_{n=0}^N$, where $\tilde{x}_n=x_n$ for $n\in\{1,2,\ldots,N\}$ and $\tilde{x}_0=0$.
	\item\label{step:max2} Find a $\tilde{x}$, which, after its transfer, minimizes the absolute difference between $S_1$ and $S_2$ (for multiple minimizers the smaller indexed one is selected), i.e., 
	\begin{align}
	\tilde{x}_*=\argmin_{\tilde{x}\in\widetilde{\mathcal{X}}}\abs{(S_1-\tilde{x})-(S_2+\tilde{x})}.\label{max2}
	\end{align}
	\item Then, move the element $\tilde{x}_*$, and update $S_1$ and $S_2$ accordingly, i.e.,
	\begin{align}
	S_1\leftarrow S_1-\tilde{x}_*,\\
	S_2\leftarrow S_2+\tilde{x}_*.
	\end{align}
	\item Then, we change the sign of $\tilde{x}_*$ to represent the transfer
	\begin{align}
	\tilde{x}_*\leftarrow-\tilde{x}_*.
	\end{align}
	\item We stop if $\tilde{x}_*=0=\tilde{x}_0$, or else, we return to Step \ref{step:max2}.
\end{enumerate}

As can be seen, we restructured the algorithm by adding a dummy zero element to the input set and altering Step \ref{step:max2} for improved performance. We stop whenever we can no longer decrease the absolute difference between the set sums.

\begin{proposition}\label{thm:S1-S2dec}
	While the algorithm continues to run, the absolute difference between the set sums strictly decreases.
	\begin{proof}
		From Step \ref{step:max2} in the algorithm, we have
		\begin{align}
			\abs{S_1-S_2-2\tilde{x}_*}\leq\abs{S_1-S_2}.
		\end{align}
		However, we stop when $\tilde{x}_*=0$, which is the only case when it holds with equality, which concludes the proof.
	\end{proof}
\end{proposition}

\begin{proposition}\label{thm:x*<S1-S2}
	During the algorithm's run, $\abs{\tilde{x}_*}<\abs{S_1-S_2}$.
	\begin{proof}
		From Step \ref{step:max2} and $0\in\widetilde{\mathcal{X}}$, we have the result.
	\end{proof}
\end{proposition}

\begin{lemma}\label{thm:once}
	Each element in $\widetilde{\mathcal{X}}$ is transferred at most once.
	\begin{proof}
		Without loss of generality, at some iteration assume that $S_1>S_2$. Let $\tilde{x}_{*,0}$, $\tilde{x}_{*,1}$ be the sequentially transfered elements. We can either have
		\begin{enumerate}
			\item $S_1-S_2-2\tilde{x}_{*,0}\geq0$, $S_1-S_2-2\tilde{x}_{*,0}-2\tilde{x}_{*,1}\geq0$;
			\item $S_1-S_2-2\tilde{x}_{*,0}\geq0$, $S_1-S_2-2\tilde{x}_{*,0}-2\tilde{x}_{*,1}\leq0$;
			\item $S_1-S_2-2\tilde{x}_{*,0}\leq0$, $S_1-S_2-2\tilde{x}_{*,0}-2\tilde{x}_{*,1}\leq0$;
			\item $S_1-S_2-2\tilde{x}_{*,0}\leq0$, $S_1-S_2-2\tilde{x}_{*,0}-2\tilde{x}_{*,1}\geq0$;
		\end{enumerate}
		
		Because of the minimization in Step \ref{step:max2}, we see that for the first two cases, we have $\abs{\tilde{x}_{*,1}}\leq \abs{\tilde{x}_{*,0}}$. For the third case, we have $\abs{\tilde{x}_{*,1}}< \abs{\tilde{x}_{*,0}}$ from \autoref{thm:x*<S1-S2}. For the fourth case, we have $\abs{\tilde{x}_{*,1}}< \abs{\tilde{x}_{*,0}}$ from \autoref{thm:S1-S2dec}. We can transfer $\abs{\tilde{x}_{*,0}}$ back only if the sign of sum difference changes at a time after its first transfer. However, because of the strict inequalities, we can never transfer it back. 
	\end{proof}
\end{lemma}

From \autoref{thm:once}, we make at most $O(N)$ transfers. Since the minimization step takes at most $O(\log N)$ time, the run time of the algorithm is $O(N\log N)$ at the worst case. Similarly, we also again have $O(N)$ space complexity.

\begin{example}
	The same working example is as follows. Let $\mathcal{X}=\{2,3,5,7,11,13,17,19,23,29\}$. We sequentially have:
	\begin{align*}
	\widetilde{\mathcal{X}}=&\{0,2,3,5,7,11,13,17,19,23,29\}, &&S_1-S_2=129,\\
	\widetilde{\mathcal{X}}=&\{-29,0,2,3,5,7,11,13,17,19,23\}, &&S_1-S_2=71,\\
	\widetilde{\mathcal{X}}=&\{-29,-23,0,2,3,5,7,11,13,17,19\}, &&S_1-S_2=25,\\
	\widetilde{\mathcal{X}}=&\{-29,-23,-13,0,2,3,5,7,11,17,19\}, &&S_1-S_2=-1.
	\end{align*}
	Hence, the resulting partition is
	\begin{align*}
	\mathcal{X}_1=\{2,3,5,7,11,17,19\} &&\mathcal{X}_2=\{13,23,29\},
	\end{align*} 
	which is not only locally optimal, but also globally optimal for this specific example.
\end{example}.
\section{Extension to Negative Set Elements}\label{sec:ext}
First of all, if $\mathcal{X}$ contains zero elements, it is inconsequential since they can be arbitrarily assigned to $\mathcal{X}_1$ and $\mathcal{X}_2$ at the end. Secondly, if the sample set includes not only positive samples but also negative ones, we can deal with them straightforwardly with the following changes to the algorithm:
\begin{itemize}
	\item Create the set $\widetilde{\mathcal{X}}=\{\tilde{x}_n\}_{n=0}^N$, where $\tilde{x}_n=\abs{x_n}$ for $n\in\{1,2,\ldots,N\}$ and $\tilde{x}_0=0$, than sort $\widetilde{\mathcal{X}}$.
	\item Create the set $\mathcal{B}=\{b_n\}_{n=1}^N$, where $b_n=\sign{x_n}$
	\item At the end of the algorithm, multiply $\widetilde{\mathcal{X}}$ with $\mathcal{B}$, i.e., 
	\begin{align}
		\tilde{x}_n\leftarrow b_n\tilde{x}_n,
	\end{align}
	then, assign the positives to $\mathcal{X}_1$ and negatives to $\mathcal{X}_2$.
\end{itemize}
\begin{example}
	An example is as follows. Let $\mathcal{X}=\{-23,-17,-11,-5,-2,3,7,13,19,29\}$. The algorithm works as follows:
	\begin{align*}
	\mathcal{B}=&\{-1,1,-1,1,-1,1,-1,1,-1,1\}&&\\
	\widetilde{\mathcal{X}}=&\{2,3,5,7,11,13,17,19,23,29\}, &&S_1-S_2=129,\\
	\widetilde{\mathcal{X}}=&\{-29,2,3,5,7,11,13,17,19,23\}, &&S_1-S_2=71,\\
	\widetilde{\mathcal{X}}=&\{-29,-23,2,3,5,7,11,13,17,19\}, &&S_1-S_2=25,\\
	\widetilde{\mathcal{X}}=&\{-29,-23,-13,2,3,5,7,11,17,19\},&&S_1-S_2=-1,
	\end{align*}
	Hence, the resulting partition is
	\begin{align*}
	\mathcal{X}_1=\{-23,3,7,19\} &&\mathcal{X}_2=\{-17,-11,-5,-2,13,29\},
	\end{align*} 
	which is not only locally optimal, but also globally optimal for this specific example.
\end{example}.

\section{Conclusion}\label{sec:conc}
We studied the optimization version of the set partition problem where the difference between the partition sums are minimized. While the set partitioning problem is NP-hard and requires exponential complexity to solve; we formulated a weaker version of this NP-hard problem, where the goal is to find a locally optimal solution. We proposed algorithms that can find locally optimal solutions in $O(N\log N)$ time and $O(N)$ space. Our algorithms require neither positive nor integer elements in the input set, hence, they are more widely applicable.

\bibliographystyle{IEEEtran}
\bibliography{double_bib}

\end{document}